\documentclass[12pt]{article}
\usepackage[english,german,french,polish]{babel}
\usepackage[T1]{fontenc}
\usepackage{amsfonts}

\begin{document}

\selectlanguage{english}

\baselineskip 0.76cm
\topmargin -0.6in
\oddsidemargin -0.1in

\let\ni=\noindent

\renewcommand{\thefootnote}{\fnsymbol{footnote}}

\newcommand{\SM}{Standard Model }

\pagestyle {plain}

\setcounter{page}{1}

\pagestyle{empty}

~~~

\begin{flushright}
IFT-- 05/18
\end{flushright}

\vspace{1.0cm}

{\large\centerline{\bf Excellent approximate solution to the}} 

{\large\centerline{\bf mysterious mass equation by Koide}} 

\vspace{0.4cm}

{\centerline {\sc Wojciech Kr\'{o}likowski}}

\vspace{0.3cm}

{\centerline {\it Institute of Theoretical Physics, Warsaw University}}

{\centerline {\it Ho\.{z}a 69,~~PL--00--681 Warszawa, ~Poland}}

\vspace{0.6cm}

{\centerline{\bf Abstract}}

\vspace{0.2cm}

It is shown that the efficient mass formula we found for charged leptons in 1992 can be considered as an excellent approximate solution to the mysterious charged-lepton mass equation proposed by Koide in 1981.

\vspace{1.0cm}

\ni PACS numbers: 12.90. +b 

\vspace{1.0cm}

\ni July 2005  

\vfill\eject

~~~
\pagestyle {plain}

\setcounter{page}{1}

\vspace{0.2cm}

In 1992 we found an efficient empirical mass formula for charged leptons $e_i = e^-, \mu^-, \tau^-$ [1]. It has the form

%rownanie 1
\begin{equation}
m_{e_i} = \mu  \rho_i  \left(N^2_i + \frac{\varepsilon  - 1}{N^2_i} \right) \,,
\end{equation}

\ni where 

%rownanie 2
\begin{equation}
N_i = 1,3,5 
\end{equation}

\ni and

%rownanie 3
\begin{equation}
\rho_i = \frac{1}{29} \,,\,\frac{4}{29} \,,\,\frac{24}{29} 
\end{equation}

\ni ($\sum_i \rho_i = 1$). Here, $\mu  > 0$ and $\varepsilon  > 0$ are two constants. With the experimental values $m_e = 0.5109989$ MeV and $m_\mu = 105.65837$ MeV as an input, the formula (1), rewritten explicitly as

%rownanie 4
\begin{equation}
m_e = \frac{\mu }{29} \varepsilon  \;, \nonumber \\  m_\mu = \frac{\mu }{29} \frac{4}{9} (80 +  \varepsilon ) \;,\;  m_\tau = \frac{\mu }{29} \frac{24}{25} (624 + \varepsilon )\;,
\end{equation}

\ni gives  the prediction

%rownanie 5
\begin{equation}
m_\tau = \frac{6}{125}(351 m_\mu - 136 m_e) = 1776.7964\;{\rm MeV} = 1776.80\;{\rm MeV}
\end{equation}

\ni that is really close to the experimental value $m_\tau = 1776.99^{+0.29}_{-0.26}\;{\rm MeV}$ [2]. The formula (1) determines also both constants

%rownanie 6
\begin{eqnarray}
\mu  & = & \frac{29(9m_\mu - 4m_e)}{320} = 85.992356\;{\rm MeV} = 85.9924 \;{\rm MeV} \;,\nonumber \\ \varepsilon & = & \frac{320 m_e}{9m_\mu - 4m_e} = 0.1723289 = 0.172329 \,.
\end{eqnarray}

Although the formula (1) has essentially an empirical character, there exists a speculative background for it [1] related to a K\"{a}hler-like extension of the Dirac equation ({\it i.e.}, the extended Dirac's square-root procedure) and to the Pauli principle realized in an intrinsic way for additional bispinor indices appearing in such an extended Dirac equation. Their number is equal to $N_i -1 = 0,2,4$, where $N_i$ denotes the total number of bispinor indices as is given in Eq. (2). This Pauli principle restricts the number of lepton and quark generations to three and only three. Then, the generation-weighting factors $\rho_i$ in the mass formula (1) are defined as in Eq. (3).

Now, consider the following ratio involving the experimental values of $m_e$ and $m_\mu $ and the predicted value (5) of $m_\tau $:

%rownanie 7
\begin{equation}
\frac{m_e + m_\mu + m_\tau}{(\sqrt{m_e} + \sqrt{m_\mu} + \sqrt{m_\tau})^2} = 0.6666569 = (1/1.0000146)\frac{2}{3} \;.
\end{equation}

\ni This appears to be very close to 2/3.

We can see that our charged-lepton mass formula (1) gives an excellent approximate solution to the mysterious mass equation proposed for charged leptons by Koide in 1981 [3]:

%rownanie 8
\begin{equation} 
\frac{m_e + m_\mu + m_\tau}{(\sqrt{m_e} + \sqrt{m_\mu} + \sqrt{m_\tau})^2} = \frac{2}{3} 
\end{equation}

\ni that, in general, leads to two different predictions for $m_\tau $, when the experimental values $m_e $ and $m_\mu $ are used as an input, namely

%rownanie 9
\begin{eqnarray}
m_\tau & = & \left[ 2\left( \sqrt{m_e}+ \sqrt{m_\mu} \right) \pm \sqrt{3(m_e + m_\mu) + 12\sqrt{m_e m_\mu}}\, \right]^2\; {\rm MeV} \nonumber \\ 
& = & \left\{\begin{array}{l} 1776.9689\; {\rm MeV}\\ 3.3173557\; {\rm MeV} \end{array} \right. = \left\{\begin{array}{l} 1776.97\; {\rm MeV}\\ 3.31736\; {\rm MeV} \end{array} \right.  \,. 
\end{eqnarray}

\ni The large solution is excellent, being even closer to the experimental value of $m_\tau $ than our prediction (5), while the small solution does not correspond to any known experimental object. This wrong solution, however, cannot be excluded from the Koide equation (8) by itself, unless it is required that $m_\mu <m_\tau $. In the case of our mass formula (1) leading to the linear mass relation (5) there is only a unique solution and it appears as really close to the experimental value of $m_\tau $ (though it is a bit worse than the large Koide solution (9)).

I am indebted to Alejandro Rivero for his kindly calling my attention to the mysterious mass equation proposed by Koide.

\vfill\eject

~~~~
\vspace{0.5cm}

{\centerline{\bf References}}

\vspace{0.5cm}

{\everypar={\hangindent=0.6truecm}
\parindent=0pt\frenchspacing

{\everypar={\hangindent=0.6truecm}
\parindent=0pt\frenchspacing

[1]~For a recent presentation {\it cf.} W. Kr\'{o}likowski, {\it Acta Phys. Pol.} {\bf B 33}, 2559 (2002) [{\tt hep--ph/0203107}], and notice the references therein; {\it cf.} also W. Kr\'{o}likowski, {\it Acta Phys. Pol.} {\bf B 36}, 2051 (2005)[{\tt hep--ph/0503074}], and {\tt hep--ph/0504256}. 

\vspace{0.2cm}

[2]~Particle Data Group, {\it Review of Particle Physics, Phys.~Lett.} {\bf B 592} (2004).

\vspace{0.2cm}

[3]~For a recent discussion {\it cf. } Y.~Koide, {\tt hep--ph/0506247}, and references therein;  {\it cf. } also A.~Rivero and A.~Gsponer, {\tt hep--ph/0505220}.

\vspace{0.2cm}

\vfill\eject

\end{document}